\documentclass[useAMS,usenatbib]{mn2e}
\usepackage{graphicx}
\usepackage{xspace}

\newcommand{\bb}[1]{\ifmmode \mbox{\boldmath $ #1$} \else  \mbox{\boldmath $#1$} \fi}
\newcommand{\bi}[1]{\bmath{#1}}
\newcommand{\dd}{\ensuremath{\,\mathrm{d}}}
\newcommand{\ddd}{\ensuremath{\mathrm{d}}}

\newcommand{\U}[1]{\ensuremath{\mathrm{~#1}}}

\newcommand{\yr}{\U{yr}}
\newcommand{\Myr}{\U{Myr}}

\newcommand{\pc}{\U{pc}}
\newcommand{\kpc}{\U{kpc}}

\newcommand{\msun}{\U{M}_{\odot}}

\newcommand{\mpc}{\U{M_{\odot}\ pc^{-3}}}

\newcommand{\br}{\bi{r}}

\newcommand{\rh}{r_{\rm h}}
\newcommand{\rp}{r_{\rm 0}}
\newcommand{\rpc}{r_{\rm 0,c}}
\newcommand{\rt}{r_{\rm t}}
\newcommand{\rg}{R_{\rm G}}
\newcommand{\rv}{r_{\rm v}}
\newcommand{\mg}{M_{\rm G}}
\newcommand{\mc}{M_{\rm c}}
\newcommand{\mdot}{\dot{\mc}}
\newcommand{\phig}{\phi_{\rm G}}
\newcommand{\phic}{\phi_{\rm c}}
\newcommand{\phie}{\phi_{\rm e}}
\newcommand{\ej}{E_{\rm J}}
\newcommand{\ecluster}{E_{\rm c}}
\newcommand{\rhog}{\rho_{\rm G}}

\newcommand{\tesc}{t_{\rm esc}}
\newcommand{\trh}{t_{\rm rh}}
\newcommand{\tdiss}{t_{\rm diss}}
\newcommand{\torb}{t_{\rm orb}}
\newcommand{\tcr}{t_{\rm cr}}
\newcommand{\tidaltensor}{\textbf{T}_{\rm t}}
\newcommand{\efftensor}{\textbf{T}_{\rm e}}
\newcommand{\lambeone}{\lambda_{\rm e,1}}
\newcommand{\lambetwo}{\lambda_{\rm e,2}}
\newcommand{\lambethr}{\lambda_{\rm e,3}}

\newcommand{\nbody}{\texttt{NBODY6}\xspace}
\newcommand{\nbodytt}{\texttt{NBODY6tt}\xspace}
\newcommand{\rhomu}{\rho_0}

\newcommand{\eqn}[2][]{Equation#1~\ref{eqn:#2}} 
\newcommand{\fig}[2][]{Figure#1~\ref{fig:#2}}
\newcommand{\tab}[2][]{Table#1~\ref{tab:#2}}
\renewcommand{\eqn}[2][]{equation#1~(\ref{eqn:#2})}
\renewcommand{\fig}[2][]{Fig#1.~\ref{fig:#2}}

\usepackage{ifthen}
\def\draftversion{1} 

\ifthenelse{\equal{\draftversion}{1}}{
	\usepackage{xcolor}
	\newcommand{\tmp}{}
	\newenvironment{envcomm}[1]{\renewcommand{\tmp}{#1}\begin{color}{blue}\begin{center}\hrule\vspace{0.5mm}\tmp's COMMENTS\end{center}}{\begin{center}END OF \tmp's COMMENTS\vspace{0.5mm}\hrule\end{center}\end{color}}
	\newenvironment{draft}{\begin{color}[rgb]{0,0.4,0}\begin{center}\hrule\vspace{0.5mm}DRAFT\end{center}}{\begin{center}END OF DRAFT\vspace{0.5mm}\hrule\end{center}\end{color}}
	\newcommand{\comcomm}[2]{\begin{color}{blue}\ $\bullet$ \textbf{#1:} #2 $\bullet$\ \end{color}}
	\newcommand{\revend}[1]{\par\begin{color}[rgb]{0,0.4,0}\begin{center}\hrule\vspace{0.5mm}END OF #1's REVISIONS\vspace{0.5mm}\hrule\end{center}\end{color}\par}
	\newcommand{\todo}[1]{\begin{color}{red}\ $\bullet$ \textbf{To do: }#1 $\bullet$\ \end{color}}
	
	}{
	\newsavebox{\trashcan}
	\newenvironment{envcomm}[1]{\begin{lrbox}{\trashcan}\begin{minipage}{\columnwidth}}{\end{minipage}\end{lrbox}}
	
	\newcommand{\comcomm}[2]{}
	\newcommand{\revend}[1]{}
	\newcommand{\todo}[1]{}
	
	}

\title{Evolution of star clusters in arbitrary tidal fields}

\author[Renaud, Gieles \& Boily]{Florent~Renaud$^{1,2,3}$\thanks{florent.renaud@cea.fr}, Mark~Gieles$^{2}$ and Christian~M.~Boily$^{1}$\\
 $^1$ Observatoire Astronomique and CNRS UMR 7550, Universit\'e de Strasbourg, 11 rue de l'Universit\'e, F-67000 Strasbourg, France\\
 $^2$ Institute of Astronomy, University of Cambridge, Madingley Road, Cambridge, CB3 0HA, UK\\
 $^3$ Laboratoire AIM Paris-Saclay, CEA/IRFU/SAp, Universit\'e Paris Diderot, F-91191 Gif-sur-Yvette Cedex, France
}
\begin{document}
\maketitle

\begin{abstract}
We present a novel and flexible tensor approach to computing the effect of a time-dependent tidal field acting on a stellar system. The tidal forces are recovered from the tensor by polynomial interpolation in time. The method has been implemented in a direct-summation stellar dynamics integrator (\nbody) and test-proved through a set of reference calculations: heating, dissolution time and structural evolution of model star clusters are all recovered accurately. The tensor method is applicable to arbitrary configurations, including the important situation where the background potential is a strong function of time. This opens up new perspectives in stellar population studies reaching to the formation epoch of the host galaxy or galaxy cluster, as well as for star-burst events taking place during the merger of large galaxies. A pilot application to a star cluster in the merging galaxies NGC~4038/39 (the Antennae) is presented.
\end{abstract}

\section{Introduction}

The halos of nearly all galaxies are populated by old globular clusters that presumably formed in gaseous discs at high redshift \citep[$z\sim3-5$,][] {2005ApJ...623..650K}. Young `populous clusters', or `super-star clusters' are found in the Large Magellanic Cloud \citep{1961ApJ...133..413H}, starburst galaxies \citep[e.g.][]{1971A&A....12..474V}, interacting galaxies and merger remnants \citep[e.g.][]{1995AJ....109..960W,1997AJ....114.2381M} and also in quiescent spirals \citep{2002AJ....124.1393L, 2008IAUS..250..247F}. This suggests that globular cluster formation is not unique to the early Universe and that the formation of these dense stellar systems is a common phenomenon in star formation \citep*{1997ApJ...480..235E, 2010ARA&A..48..431P}. 

There is increasing evidence from studies of the Milky Way and the Andromeda galaxy that some globular clusters have only recently (past $\sim\,$Gyr) been brought in through satellite accretion \citep[e.g.][]{2009ApJ...694.1498M,2010ApJ...717L..11M}. In order to understand the relation between the young massive clusters and the old globular clusters, i.e. their life-cycle, we need to place their evolution in a cosmological context. During the formation process of galaxies through repeated accretion phases, substructures such as star clusters or dwarf satellites evolve along complex orbits in a non-static external potential. This makes their evolution difficult (if not impossible) to describe analytically.

To approach this problem numerically is also challenging because of the large range of evolutionary time scales involved, ranging from several days for tight binaries in the cores of clusters to a Hubble time for the host galaxy \citep[see the recent review by][]{2011EPJP..126...55D}. To be able to self-consistently model the evolution of star clusters in `live' galaxies one needs to rely on a direct-tree hybrid approach  \citep[e.g.][]{2007PASJ...59.1095F}.

This is why most studies of the (dynamical) evolution of star clusters simplify the effect of the external tidal field by assuming a static background potential \citep[e.g.][]{1990ApJ...351..121C, 1997MNRAS.289..898V,2001ApJ...561..751F, 2003MNRAS.340..227B,2004AJ....127.2753D,2008MNRAS.389L..28G,2008AJ....135.2129H,2009MNRAS.399.1275P,2011MNRAS.411.1989Z}. Although this is probably an adequate approximation for many purposes, it does not suffice for more complicated orbits such as those of clusters in mergers of massive galaxies, or satellite galaxies that are accreted.
 
In light of the last point, several recent studies have adopted a (semi-)analytical approach to star cluster evolution in more realistic external tides, such as the hierarchical build-up of galaxies \citep[e.g.][]{2008ApJ...689..919P} and galaxy mergers \citep[e.g.][]{2011MNRAS.414.1339K}. In here the effect of mass loss because of stellar evolution, evaporation of stars over the tidal radius and the shock-enhanced escape of stars because of rapidly varying tidal fields (i.e. disc crossings and bulge shocks) are applied analytically. In almost all cases these processes are assumed to be independent of each other such that the individual resulting mass-loss rates are simply added.

An important, and often dominant, mass-loss mechanism is the relaxation driven escape of stars, so-called evaporation. Some models assume that a constant fraction of the stars evaporate per half-mass relaxation time, independent of the orbit \citep[e.g.][]{1997ApJ...474..223G,2008ApJ...689..919P}. Others consider that the escape fraction depends on the galactocentric radius, often assumed to be the pericentre distance \citep{1962AJ.....67..471K,1983AJ.....88..338I,2001ApJ...561..751F}. The final lifetime of clusters can be quite different, depending on the details of the assumptions that are made.

Another critical disruptive agent is mass-loss  due to external tidal perturbations, or `shocks'. The related lifetime scales linearly with the density of the cluster \citep{1958ApJ...127...17S,1972ApJ...176L..51O} and it is, therefore, that the results of semi-analytical models rely critically on what is assumed for the relation between the cluster mass and the half-mass radius; a relation which is not only affected by evaporation (i.e. because of the reduction of the mass in time), but also because of relaxation driven expansion \citep{1965AnAp...28...62H, 1984ApJ...280..298G,2002MNRAS.336.1069B,2010MNRAS.408L..16G}, which as far as we are aware is neglected in all semi-analytic approaches. For clusters on circular orbits in static potentials that are well within the tidal limit initially (the half-mass radius being less than a few percent of the tidal radius), it was found that the expansion phase dominates the evolution in the first half of the cluster's lifetime, while evaporation dominates in the second half \citep{2011MNRAS.413.2509G}. In the latter stage the cluster density adjusts to the mean (galactic) density along the orbit \citep{2010MNRAS.407.2241K}. It is, therefore, necessary to consider both the internal evolution (relaxation) and the external effects (tides) simultaneously if one considers the entire life-cycle of clusters.

Current direct $N$-body codes are capable of solving the $N$-body problem numerically for $N\sim10^5$, together with the effects of mass-loss of the individual stars, binary interactions and tidal fields \citep{2001MNRAS.321..199P,2003gnbs.book.....A}. Thanks to increasing computational power, it is now possible to combine the relaxation driven evolution of star clusters in cosmologically motivated external conditions. In this paper we present a new method that integrates the effect of any tidal field to the evolution of galaxy substructures, that we specifically apply to star clusters. We do this by extracting the tidal tensor, that contains all information about the tidal field at the location of the substructure, from galaxy simulations and subsequently `feed' this to a stellar dynamics code. However, we have not yet implemented the other half of the scale coupling, as the feedback from the small scales (e.g. metal enrichment, stellar winds, supernova explosions) is not retroceded to the ambient galactic medium.

The paper is organised as follows: first, we setup the framework for the computation of the tidal acceleration by means of tensors (Section~\ref{sec:generalcase}). The expressions found are then applied to the special cases of circular orbits: well-known expressions are retrieved using the new formalism in Section~\ref{sec:circular}. The role of the galactic profile on the evolution of clusters through the escape of stars is specially explored in Section~\ref{sec:escape}. The limitations of the analytical approach are explained in Section~\ref{sec:noncircular}, while Section~\ref{sec:nbody6tt} presents the numerical implementation of the method to compute the tidal forces in a stellar dynamics code. A comparison with previous results obtained for idealized configurations is carried out. Applications to innovative cases are presented as the first practical illustrations of the method. Finally, the limitations and some possible future developments of our approach are discussed.

\section{Analytical description of arbitrary tidal fields}
\label{sec:generalcase}

The main goal of the paper is to provide a general framework within which to follow the evolution of self-gravitating stellar associations in arbitrary and time-dependent tidal fields. For concreteness in the remainder of the paper we focus on star clusters orbiting within a galaxy in equilibrium, but the formalism can be exported to many other situations (such as dwarfs galaxies, galaxy mergers, galaxy clusters, etc).

\subsection{Tidal and effective tensors}
\label{sec:tensors}

It is convenient to work in coordinates centred on the star cluster being embedded in the background gravitational potential, as opposed to the global system's barycentre. The tides derive from gradients in the external gravitational acceleration across the cluster. Subtracting the acceleration of the cluster's centre of mass by the host galaxy, the relative acceleration of a member star at the position $\br'$ in this frame reads
\begin{equation} 
\frac{\ddd^2 \br'}{\ddd t^2} = - \bb{\nabla}\phic(\br') - \bb{\nabla}\phig(\br') + \bb{\nabla}\phig(\bb{0}),
\label{eqn:gravacceleration}
\end{equation}
where $\phic$ and $\phig$ are the gravitational potentials of the star cluster and the host galaxy, respectively. The standard treatment of the background gravity in the tidal limit \citep[e.g.][]{2008gady.book.....B} consists in regrouping the last two terms in \eqn{gravacceleration} and performing a linear expansion by considering that $r' \ll \rg$, with $\rg$ the distance between the cluster and the galaxy's barycentre. It is important to recall that the linear expansion will hold when and if the Laplacian of the galaxy's potential is small at the location of the cluster, regardless of the ratio $r' / \rg$. Bearing this in mind, the linearised equations of motion are expressed in general form through the tidal tensor $\tidaltensor$ of components
\begin{equation}
T_\mathrm{t}^{ij}(\br') = \left(-\frac{\partial^2 \phig}{\partial {x'}^i \ \partial {x'}^j}\right)_{\br'}.
\label{eqn:tt}
\end{equation}
To first order in $\br'$ we have
\begin{equation}
\bb{\nabla}\phig(\br') = \bb{\nabla}\phig(\bb{0}) - \tidaltensor(\br') \cdot \br' + {\cal O}(\br'^2).
\end{equation}
Substituting in \eqn{gravacceleration}:
\begin{equation}
\frac{\ddd^2 \br'}{\ddd t^2} = - \bb{\nabla}\phic(\br') + \tidaltensor(\br') \cdot \br'.
\end{equation}
The symmetry $T_\mathrm{t}^{ij} = T_\mathrm{t}^{ji}$ allows us to express $\tidaltensor$ in diagonal form in the base of its eigenvectors $\bb{\nu}_i$, ($i = 1$ to 3): the amplitude of the eigenvalues $\lambda_i$ is a measure of the strength of the tidal field along the corresponding eigenvector. When the proper base of the tensor is used to express the accelerations, the reference frame becomes non-inertial. Even so, only a rotational component at angular frequency $\bb{\Omega}$ appears because the translational component is absorbed in \eqn{gravacceleration}. The net acceleration now includes non-inertial terms from fictitious forces:
\begin{eqnarray}
\frac{\ddd^2 \br}{\ddd t^2} &=& \overbrace{ \overbrace{-\bb{\nabla}\phic(\br)}^{\textrm{internal}}\ \overbrace{ +\tidaltensor(\br)\cdot\br }^{\textrm{tidal}}}^{\textrm{gravitational}} \nonumber \\
&& \underbrace{ \underbrace{\phantom{\frac{}{}}- \bb{\Omega}\times\left(\bb{\Omega}\times \br\right)}_{\textrm{centrifugal}}\ \underbrace{- \frac{\ddd\bb{\Omega}}{\ddd t} \times \br}_{\textrm{Euler}}\ \underbrace{-2\bb{\Omega}\times \frac{\ddd\br}{\ddd t}}_{\textrm{Coriolis}} }_{\textrm{fictitious}},
\label{eqn:allacceleration}
\end{eqnarray}
where $\br$ is the position vector in the non-inertial frame. The centrifugal acceleration can be derived from the gradient of a scalar potential
\begin{equation}
\phi_{\rm f}(\br) = \frac{1}{2} \left(\br\cdot\bb{\Omega}\right)^2 - \frac{1}{2} \bb{\Omega}^2\bb{r}^2,
\end{equation}
defined up to an arbitrary additive constant. This in turn leads to an effective tidal potential $\phie$ and the associated effective tidal tensor $\efftensor$ of components
\begin{equation}
T_{\rm e}^{ij}(\br) = T_{\rm t}^{ij}(\br) + \left(-\frac{\partial^2 \phi_{\rm f}}{\partial x^i \ \partial x^j}\right)_{\br} \equiv \left(-\frac{\partial^2 \phie}{\partial x^i \ \partial x^j}\right)_{\br}.
\label{eqn:efft}
\end{equation} 
The total acceleration becomes
\begin{equation}
\frac{\ddd^2 \br}{\ddd t^2} = -\bb{\nabla}\phic(\br) + \efftensor(\br) \cdot \br - \frac{\ddd\bb{\Omega}}{\ddd t} \times \br - 2 \bb{\Omega}\times \frac{\ddd \br}{\ddd t}.
\label{eqn:effectiveacceleration}
\end{equation}
In diagonal form, we write the effective tensor $\efftensor$ as
\begin{equation}
\efftensor(\br) = \left(\begin{array}{ccc}
\lambeone & 0 & 0\\
0 & \lambetwo & 0\\
0 & 0 & \lambethr 
\end{array}\right),
\end{equation}
with the convention $\lambeone \geq \lambetwo \geq \lambethr$. In the rest of the paper we will refer to the three $\lambda_{\rm e}$'s as the effective eigenvalues.

Equation~(\ref{eqn:effectiveacceleration}) cannot be simplified for a non-zero Euler acceleration. Hence, the following analytical Sections~\ref{sec:tidalradius}, \ref{sec:jacobisurface} and \ref{sec:circular} focus on cases where $\bb{\Omega}$ is constant in time. More general configurations will be explored numerically in Section~\ref{sec:nbody6tt}.

\subsection{Tidal radius}
\label{sec:tidalradius}

The positions where the internal gravitational acceleration of the cluster is exactly balanced by all the other accelerations are called the Lagrange points $L_i$ (with $i = 1$ to 5). By convention, $L_1$ and $L_2$ fall down the galaxy-cluster axis ($L_1$ being between the two objects). The distance between the centre of the cluster and $L_1$ is referred to as the tidal radius $\rt$. 

At $L_1$, it is reasonable to approximate the potential of the cluster with that of a point of mass $\mc$. Furthermore, the effective tidal acceleration\footnote{The eigenvector related to the largest eigenvalue $\lambeone$ points toward the galaxy.} there is $\lambeone\rt$. Finally, the Lagrange points are static in this reference frame, meaning that the Coriolis acceleration of $L_1$ is zero. With these considerations, \eqn{effectiveacceleration} gives the expression of the tidal radius:
\begin{equation}
\rt = \left(\frac{G\mc}{\lambeone}\right)^{1/3}.
\label{eqn:rt}
\end{equation}
Note that this definition applies to all galactic potentials.

The sphere of radius $\rt$ can be seen as an approximation of the physical boundary of the cluster. A more precise three-dimensional definition, called the Jacobi surface, can also be used.

\subsection{Jacobi surface}
\label{sec:jacobisurface}

The effective tidal potential derives from the linearization of \eqn{efft}:
\begin{equation}
\phie(\br) = -\frac{1}{2}\ \br^\top \cdot\efftensor(\br)\cdot\br
\label{eqn:effectiveenergy}
\end{equation}
where $\br^\top$ is the transpose vector of $\br$. Therefore, with the point mass approximation for the cluster potential, the total potential is
\begin{equation}
\phi(x,y,z) = -\frac{G\mc}{\sqrt{x^2+y^2+z^2}} -\frac{\lambeone}{2} \left(x^2 + \frac{\lambetwo}{\lambeone} y^2 + \frac{\lambethr}{\lambeone} z^2 \right).
\label{eqn:potentialenergy}
\end{equation}

The three-dimensional surface of equipotential passing in $L_1$ is called the Jacobi surface. From \eqn{potentialenergy}, we find that the corresponding potential energy (also called critical energy) is
\begin{equation}
\ej = -\frac{3}{2}\frac{G\mc}{\rt}.
\label{eqn:ej}
\end{equation}

The equality of equations~(\ref{eqn:potentialenergy}) and (\ref{eqn:ej}) defines the equation of the Jacobi surface:
\begin{equation}
0 = 2\rt^3 + \sqrt{x^2+y^2+z^2} \left(x^2 + \frac{\lambetwo}{\lambeone} y^2 + \frac{\lambethr}{\lambeone} z^2 - 3 \rt^2 \right).
\label{eqn:jacobisurface}
\end{equation}

\begin{figure}
\includegraphics{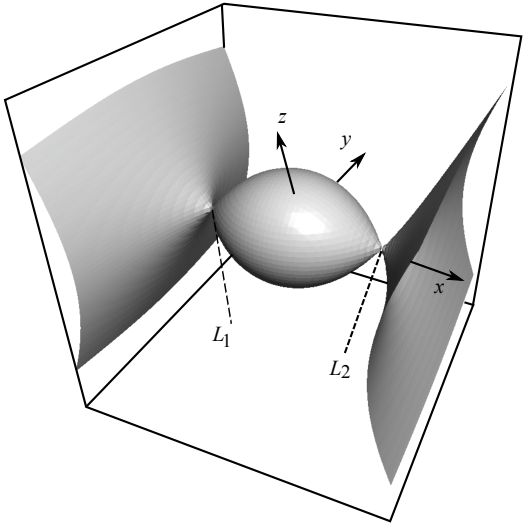}
\caption{Jacobi surface computed from \eqn{jacobisurface} for a cluster in circular orbit around a point-mass galaxy (equation~\ref{eqn:lambdaspt}).}
\label{fig:jacobisurface}
\end{figure}

A star whose energy is exactly $\ej$ cannot pass through this surface, and thus can only escape through the points $L_1$ or $L_2$, where the surface is `opened' (see an example in \fig{jacobisurface}). At energies higher than $\ej$, the apertures of the surface are larger. This plays a non-trivial role in the escape rate, as discussed in Section~\ref{sec:escape}.

\section{Application to circular orbits}
\label{sec:circular}

In this Section, we apply the formulae obtained in Section~\ref{sec:generalcase} to the special case of circular orbits around various galaxies sitting in $(-\rg, 0,0)$. The centrifugal term is derived from the rotation speed of the co-rotating reference frame which is, by definition, the orbital angular velocity $\bb{\Omega} = (0,0,\Omega)$:
\begin{equation}
\Omega = \sqrt{\frac{G\mg(\rg)}{\rg^3}},
\label{eqn:omega_pt}
\end{equation}
where $\mg(\rg)$ is the mass of the galaxy enclosed within the orbital radius $\rg$, and $\mg$ (with no argument) is the total mass of the galaxy, and $\mc \ll \mg(\rg)$.

The centrifugal acceleration is strictly opposed to the tidal contribution along the $y$- and $z$-axes:
\begin{equation}
\Omega^2 \br = -\left(-\frac{\partial^2 \phig}{\partial y^2}\right)_{\rg} \br \ \ \ =\ -\left(-\frac{\partial^2 \phig}{\partial z^2}\right)_{\rg} \br,
\end{equation}
so that the effective tidal eigenvalues are
\begin{equation}
\left\{\begin{array}{l}
\displaystyle\lambeone = \left(-\frac{\partial^2 \phig}{\partial x^2}\right)_{\rg} - \left(-\frac{\partial^2 \phig}{\partial z^2}\right)_{\rg}\\
\displaystyle\lambetwo = 0 \\
\displaystyle\lambethr = \left(-\frac{\partial^2 \phig}{\partial z^2}\right)_{\rg},
\end{array}\right.
\label{eqn:eigenvaluescircular}
\end{equation}
for all circular orbits.

\subsection{Point-mass galaxy}

\begin{figure*}
\includegraphics{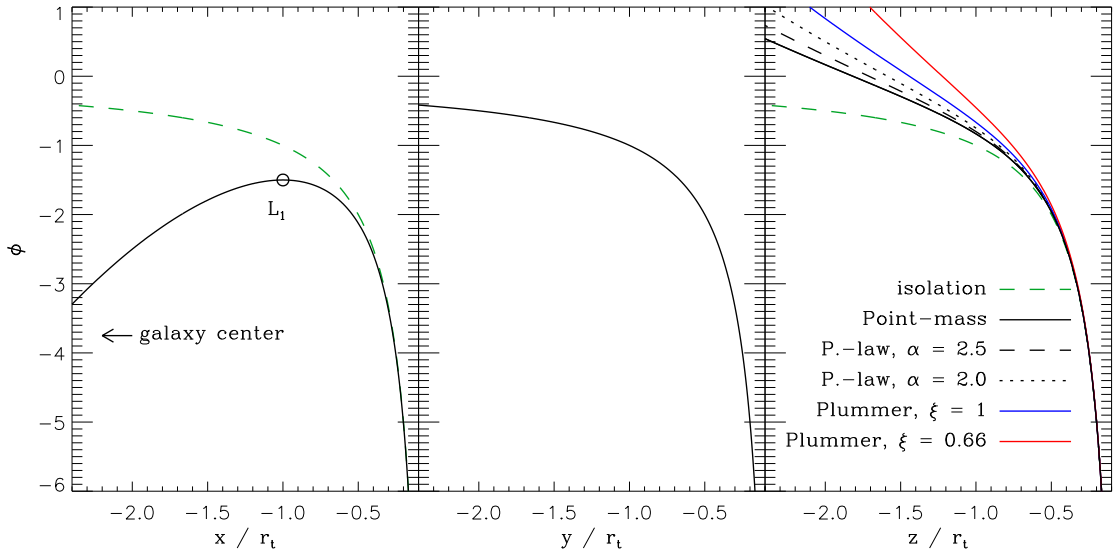}
\caption{Projections of the total potential (from equation~\ref{eqn:potentialenergy}, with $G=\mc=\rt=1$), along the $x$-, $y$- and $z$-axes, for the cluster in circular orbit in the $xy-$plane around a point-mass galaxy, galaxies with power-law density profile of slope 2.5 and 2.0, and a Plummer galaxy with the cluster in the tidally extensive ($\xi=1$) and compressive regimes ($\xi=0.66$, see text). Along the $x$- and $y$-axes, all these projections are identical. A circle marks the position of the Lagrange point $L_1$. The potential of the cluster in isolation ($r^{-1}$, green) is also plotted, for comparison.}
\label{fig:potential}
\end{figure*}

We first focus on the academic case of a cluster in circular orbit around a point-mass galaxy. Using \eqn{eigenvaluescircular}, we find that the triplet of effective eigenvalues reads
\begin{equation}
\left\{\lambeone,\lambetwo,\lambethr\right\} = \frac{G\mg}{\rg^3} \left\{3, 0, -1\right\},
\label{eqn:lambdaspt}
\end{equation}
Note that when replacing this value of $\lambeone$ in \eqn{rt}, we recover the well-known expression of the tidal radius:
\begin{equation}
\rt = \rg \left(\frac{\mc}{3\mg}\right)^{1/3} = \left(\frac{G\mc}{3\Omega^2}\right)^{1/3},
\label{eqn:rj3}
\end{equation}
(see \citealt{1962AJ.....67..471K}, \citealt{2000MNRAS.318..753F}, or \citealt{2008gady.book.....B}). The corresponding Jacobi surface is displayed in \fig{jacobisurface}, and the one-dimensional projections of the total potential (solid black) are compared to those of a cluster in isolation (dashed green) in \fig{potential}. The potential yields a saddle shape in $L_1$ (and in $L_2$, by symmetry).

\subsection{Power-law galaxy}
\label{sec:plgalaxy}

The formalism of the tidal tensor allows us to evaluate the tidal acceleration from any galactic potential. As an illustration, we now focus on power-law galactic profiles of index $\alpha<3$ whose density is
\begin{equation}
\rhog(x,y,z) = \rhomu \left[(x+\rg)^2+y^2+z^2\right]^{-\alpha/2}, 
\end{equation}
$\rhomu$ being a constant. At the position of the cluster $(0,0,0)$, the effective eigenvalues are
\begin{equation}
\left\{\lambeone,\lambetwo,\lambethr\right\} = \frac{4\pi G\rhomu}{(3-\alpha)\rg^\alpha} \left\{\alpha, 0, -1\right\},
\end{equation}
which is comparable to the point-mass case discussed above. This sets the value of the tidal radius at
\begin{equation}
\rt = \rg^{\alpha/3} \left(\frac{\mc (3-\alpha)}{4\pi\rhomu \alpha}\right)^{1/3} =  \left(\frac{G\mc}{\alpha \Omega^2}\right)^{1/3}.
\label{eqn:rtpl}
\end{equation}

The projections of the potential are plotted in \fig{potential} for $\alpha=2.5$ and 2.0. When normalized to the tidal radius, only the $z$-component differs from the point-mass case. The impact of this difference is further explored in Section~\ref{sec:escape}.

\subsection{Plummer galaxy}
\label{sec:plummergalaxy}

Consider now a \citet{1911MNRAS..71..460P} potential of characteristic radius $\rp$, once again centered on ($-\rg$,0,0),
\begin{equation}
\phig = -\frac{G\mg}{\left[\rp^2+(x+\rg)^2+y^2+z^2\right]^{1/2}}.
\end{equation}
We introduce $\xi = \rg/\rp$ and evaluate the effective eigenvalues at the position of the cluster:
\begin{equation}
\left\{\lambeone,\lambetwo,\lambethr\right\} = \frac{G\mg}{\rp^3 \left(1+\xi^2\right)^{3/2}} \left\{\frac{3\xi^2}{\left(1+\xi^2\right)}, 0,-1\right\}.
\label{eqn:plummerlambdas}
\end{equation}

First, we consider $\xi = 1$ so that the cluster lies in the tidally extensive regime of the Plummer sphere, i.e. where the tidal contribution to the first effective eigenvalue is positive \citep{2008MNRAS.391L..98R}. The triplet of effective eigenvalues is
\begin{equation}
\left\{\lambeone,\lambetwo,\lambethr\right\} = \frac{G\mg}{\rp^3} \left\{\frac{3\sqrt{2}}{8}, 0,-\frac{\sqrt{2}}{4}\right\},
\end{equation}
which gives a tidal radius\footnote{Substituting $\alpha$ in \eqn{rtpl} with the local slope of the density profile of the Plummer model (here 5/2 for $\xi=1$) does not lead to the correct tidal radius because the angular frequency $\Omega$ differs between the Plummer and the power-law galaxies.} of
\begin{equation}
\rt = \rp \left(\frac{\mc}{\frac{3\sqrt{2}}{8}\mg}\right)^{1/3} = \left(\frac{G\mc}{\frac{3}{2} \Omega^2}\right)^{1/3}.
\label{eqn:rtextensive}
\end{equation}

Second, by decreasing $\xi$, we shift the cluster toward the centre of the Plummer potential. The tidal contribution to $\lambeone$ first tends toward zero; a value reached for $\xi = 2^{-1/2}$. For smaller values of $\xi$, the cluster lies in the cored region of the galactic potential, and thus is in compressive tidal mode \citep[see][]{2009ApJ...706...67R}: the tidal acceleration acts in the same direction as the internal gravitational acceleration of the star cluster. Following \citet{1942psd..book.....C}, Appendix~\ref{app:compressive} demonstrates that the centrifugal contribution always compensates the compressive tidal acceleration on circular orbits, so that the Lagrange points still exist and thus, the tidal radius can be defined, even in compressive tidal mode.

The expression of $\lambeone$ in \eqn{plummerlambdas} reveals that a given tidal radius can be obtained at two different galactic radii: one in tidally extensive mode, and one in compressive mode. The compressive counterpart of our extensive example (equation~\ref{eqn:rtextensive}) is obtained for $\xi \approx 0.66$. Both cases are plotted in \fig{potential}. Interestingly, the potential in the extensive case ($\xi=1$) is strictly identical to that found with a power-law galaxy of index $\alpha = 1.5$: the evolution of identical star clusters set in these two configurations would be impossible to distinguish.

The comparison between these five configurations emphasizes that the potential well of a star cluster not only depends on the tidal radius, but also on the three-dimensional shape of the effective tensor, which varies from galaxy to galaxy. In the next Subsection, we examine the implication of this on the escape rate of stars from the cluster.

\subsection{Escape rate}
\label{sec:escape}

For circular orbits, the escape rate varies, to first order, as $\mdot \propto \Omega$ \cite[e.g.][]{1987ApJ...322..123L,2003MNRAS.340..227B,2008MNRAS.389L..28G}. However, the constant of proportionality contains details of the shape of the density profile of the galaxy. \citet{2010PASJ...62.1215T} recently demonstrated the importance of this secondary effect. They did this by considering clusters in galaxies with different power-law density profiles (as in our Section~\ref{sec:plgalaxy}). The constant $\rhomu$ was varied such that the tidal radius was the same in all cases. They found, somewhat counter-intuitively, that in this set-up the clusters with the \emph{lowest} orbital angular velocity $\Omega$ had the \emph{highest} escape rate. In the following, we confirm and generalize their result, using the tensor formalism.

\begin{figure}
\includegraphics[width=\columnwidth]{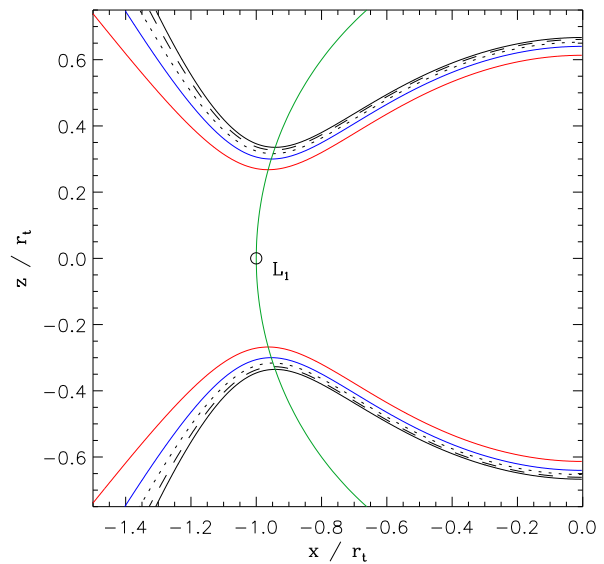}
\caption{Projection of the surface of equipotential in the $xz$-plane, for an energy $E = 0.95 \ej$ and for the five galaxies considered in \fig{potential}. The green circle represent the tidal radius.}
\label{fig:apertures}
\end{figure}

To escape from the cluster, a star needs (1) to be able to fly outside of the Jacobi surface and (2) to exit in a way not to fall back in. The first condition implies that the total energy $E = v^2/2 + \phi$ of the star must exceed the Jacobi energy: $E>\ej$. The second condition tells us that a candidate escaper which fulfills the first condition can still be trapped in the potential well of the cluster for many crossing-times \citep{2000MNRAS.318..753F, 2001MNRAS.325.1323B}: the potential barrier keeps increasing with the distance in all directions except along the galaxy-cluster axis linking $L_1$ and $L_2$ (see \fig{potential}). Therefore, the actual escapers are stars (1) with exceeding energy and (2) flying through the apertures in the corresponding equipotential surface around $L_1$ and $L_2$. The size of these apertures depends on the excess of energy and on the shape of the tidal field (see \fig{apertures}).

\begin{figure}
\includegraphics[width=\columnwidth]{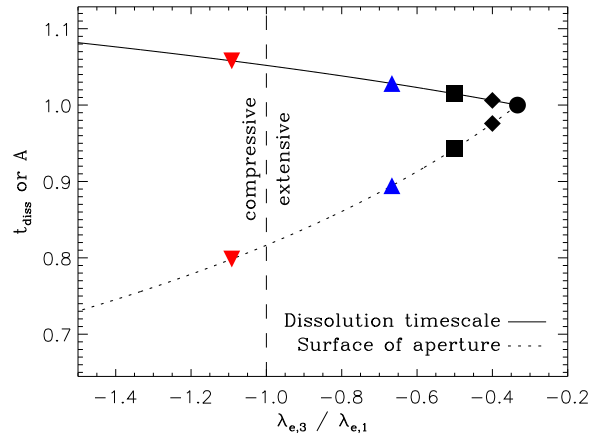}
\caption{Dissolution timescale ($\tdiss$, from equation~\ref{eqn:dissolution}) and surface of aperture in the Jacobi surface around $L_1$ and $L_2$ ($A$, computed from equation~\ref{eqn:jacobisurface}) as functions of the ratio of the effective eigenvalues, for a given tidal radius and a given cluster. Both quantities are normalized to the values they take in the case of a point-mass galaxy. The tidally compressive regime is separated from the extensive one by the vertical line. Symbols mark the cases illustrated in \fig{potential} and \ref{fig:apertures} (from left to right: Plummer $\xi = 0.66$, Plummer $\xi = 1$, power-law $\alpha = 2.0$, power-law $\alpha = 2.5$, point-mass).}
\label{fig:dissolutiontime}
\end{figure}

To evaluate the escape time, we seek the flux of stars through the apertures. We compute it by repeating the calculation applied by \citet{2000MNRAS.318..753F} to point-mass galaxies, but here for any galactic profile by means of the effective eigenvalues. Details of the derivation are given in Appendix~\ref{app:flux}. We find that the timescale for escape for a star with an excess of energy $E-\ej$ is
\begin{equation}
\tesc(E) = \frac{2}{\pi \sqrt{3}}\ C\ G\mc \frac{\sqrt{-\ej}}{(E-\ej)^2}\ \sqrt{1-\frac{\lambethr}{\lambeone}},
\label{eqn:escapetime}
\end{equation}
where $C$ is a dimensionless constant which depends on the intrinsic properties of the cluster and which we compute by numerical integration ($C \simeq 0.4$, see Appendix~\ref{app:flux}). \citet{2001MNRAS.325.1323B} wrote the (time dependent) dissolution timescale $\tdiss$ of the entire cluster as
\begin{equation}
\tdiss \propto \trh^{3/4} \tesc^{1/4}(E=2\ej),
\label{eqn:baumgardt}
\end{equation}
where $\trh$ is the half-mass relaxation time. This relation holds for homologous clusters for which the half-mass radius scales linearly with the tidal radius. We can then write the dependence of the dissolution timescale for circular orbits on the galactic parameters as:
\begin{equation}
\tdiss \propto \rt^{3/2} \left(1-\frac{\lambethr}{\lambeone}\right)^{1/8}.
\label{eqn:dissolution}
\end{equation}
We observe a first order dependence of the dissolution timescale on the tidal radius, but also a second order effect due to the shape of the galactic potential. This relation confirms and extends the conclusion of \citet{2010PASJ...62.1215T} to any galaxy: for a given tidal radius, a highly negative ratio of the effective eigenvalues corresponds to a slow dissolution, as illustrated in \fig{dissolutiontime}.

The proportionality factor in \eqn{dissolution} depends on the properties of the cluster ($\trh$, $C$, $\mc$), which all evolve with time. Thus, $\tdiss$ is an \emph{instantaneous} estimate of the dissolution timescale, and should not be mistaken with the actual life-time of the cluster. Moreover, the evolution of these properties depends on the tidal field and is very involved (if not impossible) to estimate analytically. In Section~\ref{sec:otherprofiles}, numerical experiments show indeed that the actual life-time can be very different from the analytical value of $\tdiss$ given by \eqn{dissolution}.

\section{Non-circular orbits}
\label{sec:noncircular}

As stated earlier, writing the tidal tensor in its proper base makes the reference frame non-inertial: the fictitious accelerations must be added. In the cases of linear or circular orbits, the Euler force is null and the formalism of the effective tensor allows for a simple mathematical description of the net acceleration (see Section~\ref{sec:circular}). However, for time dependent rotations, the Euler effect must be added: the effective tensor does not suffice to describe the full impact of the galaxy, and the formalism looses its advantage of analytical simplicity.

However, nothing forbids to write the tidal tensor in the inertial frame, where it is \emph{non-diagonal}, and to compute the tidal acceleration numerically. The major advantage of such a method is that the centrifugal, Euler and Coriolis accelerations, which are difficult to evaluate along complex orbits, are not required anymore. That is, the tidal tensor computed in the inertial reference frame fully represents the galactic acceleration on star clusters and allows for a numerical treatment in \emph{any} galaxy and along \emph{any} orbit. The next Section explains how this method can be implemented in an $N$-body code.

\section{$N$-body simulations: \nbodytt}
\label{sec:nbody6tt}

When expressed in the inertial reference frame, the tidal tensor contains all the information on the effect of the galaxy on its cluster. Therefore, the equations of motion of all the stars of the cluster can be solved numerically, for any tidal field. In this Section, we briefly present one implementation and a suite of tests, before applying the method to innovative cases.

\subsection{Retrieving the external force}
\label{sec:retrievingforce}

The simulations of the star clusters are done with the stellar dynamics code \nbody and its version for Graphics Processing Unit (GPU, Aarseth 2010\footnote{{\tt http://www.ast.cam.ac.uk/$\sim$sverre/web/pages/nbody.htm}}). Several cases of galactic potentials already exist in \nbody and have been widely used in previous studies. For testing purposes, we use these features and refer to them as built-in methods. On top of these pre-existing tools, we have modified \nbody to include the tidal forces by means of the tidal tensor: this new version of the code is called \nbodytt. 

Before the simulation, the tidal tensor is computed in the inertial reference frame (equation~\ref{eqn:tt}): its nine components are sampled along the orbit of the cluster within the galaxy, either analytically, or by means of independent galactic simulations \citep[see][]{2009ApJ...706...67R, 2010PhDT.........1R}. The sampling frequency is chosen to ensure that the high frequency features of the tidal acceleration, both in term of intensity and orientation, are recovered. A table of sampled tensors is then passed to \nbodytt. During the simulation of the cluster, the components of the tensor are quadratically interpolated whenever the gravitational force on a particle needs to be updated. For a star of mass $m$ at the position $\{x'^i\}$ with respect to the centre of the cluster, the tidal force, whose $i$-th component reads
\begin{equation}
F_{\rm t}^{i} = m \sum_j T_{\rm t}^{ij} x'^j,
\end{equation}
is added to the gravitational force due to the $N-1$ other particles \citep[see][for details on the solving of the $N$-body problem]{2003gnbs.book.....A}.

As soon as physically-time-dependent processes (like stellar evolution) are not involved, the entire study remains scale-free. The units adopted below are the cluster's $N$-body units \citep[$G = \mc = -4\ecluster = 1$, with $\ecluster$ being the total energy of the cluster; see][]{1986LNP...267..233H}. One of the possible scalings to physical units is proposed in \tab{units}.

\begin{table*}
\caption{One possible scaling of the simulations in physical units}
\label{tab:units} 
\begin{tabular}{ccc}
\hline
\hline
Quantity & $N$-body units & Physical units \\
(at initial time) & & \\
\hline
\multicolumn{3}{c}{Cluster scale}\\
Mass ($\mc$) & $1$ & $8 \times 10^{3} \msun$ \\
Mass of a particle ($m$) & $1/8000$ & $1 \msun$ \\
Virial radius ($\rv$) & $1$ & $1 \pc$ \\
Characteristic radius ($\rpc$) & $3\pi/16$ & $0.59 \pc$ \\
Half-mass radius ($\rh$) & $\rpc / \sqrt{2^{2/3}-1}$ & $0.77 \pc$ \\
Crossing time ($\tcr$) & $2\sqrt{2}$ & $0.47 \Myr$ \\
\hline
\multicolumn{3}{c}{Galactic scale}\\
Mass ($\mg$) & $1.25 \times 10^6$ & $10^{10} \msun$ \\
1$^{\rm st}$ effective eigenvalue$^\star$ ($\lambeone$) & $[3.75,0.13] \times 10^{-3}$ & $[135.0,5.0] \times 10^{-3} \U{Myr^{-2}}$ \\
Orbital radius$^\star$ ($\rg$) & $[1,3] \times 10^3$ & $[1,3] \kpc$ \\
Orbital period$^\dag$ ($\torb$) & $[178.3, 927.7, 504.8]$ & $[29.6, 154.0, 83.8] \Myr$ \\
\hline
\end{tabular}\\
$^\star$ for the orbits A and B, respectively.\\
$^\dag$ for the orbits A, B and C, respectively.
\end{table*}

In the inertial reference frame, it is generally not possible to evaluate the effective eigenvalues and thus, the tidal radius and the energy of the Jacobi surface. Therefore, to define the cluster membership, we have adopted an empirical criterion: a star is considered as a cluster member when the sum of its kinetic energy exactly balances the potential energy due to the $N-1$ other stars, $N$ being determined iteratively until convergence \citep[see][]{2006ApJ...645..240P}. For circular orbits, we have measured that this count only differs from the number of stars within the tidal radius by a few percent.

\subsection{Tests}
\label{sec:tests}

\begin{figure*}
\includegraphics{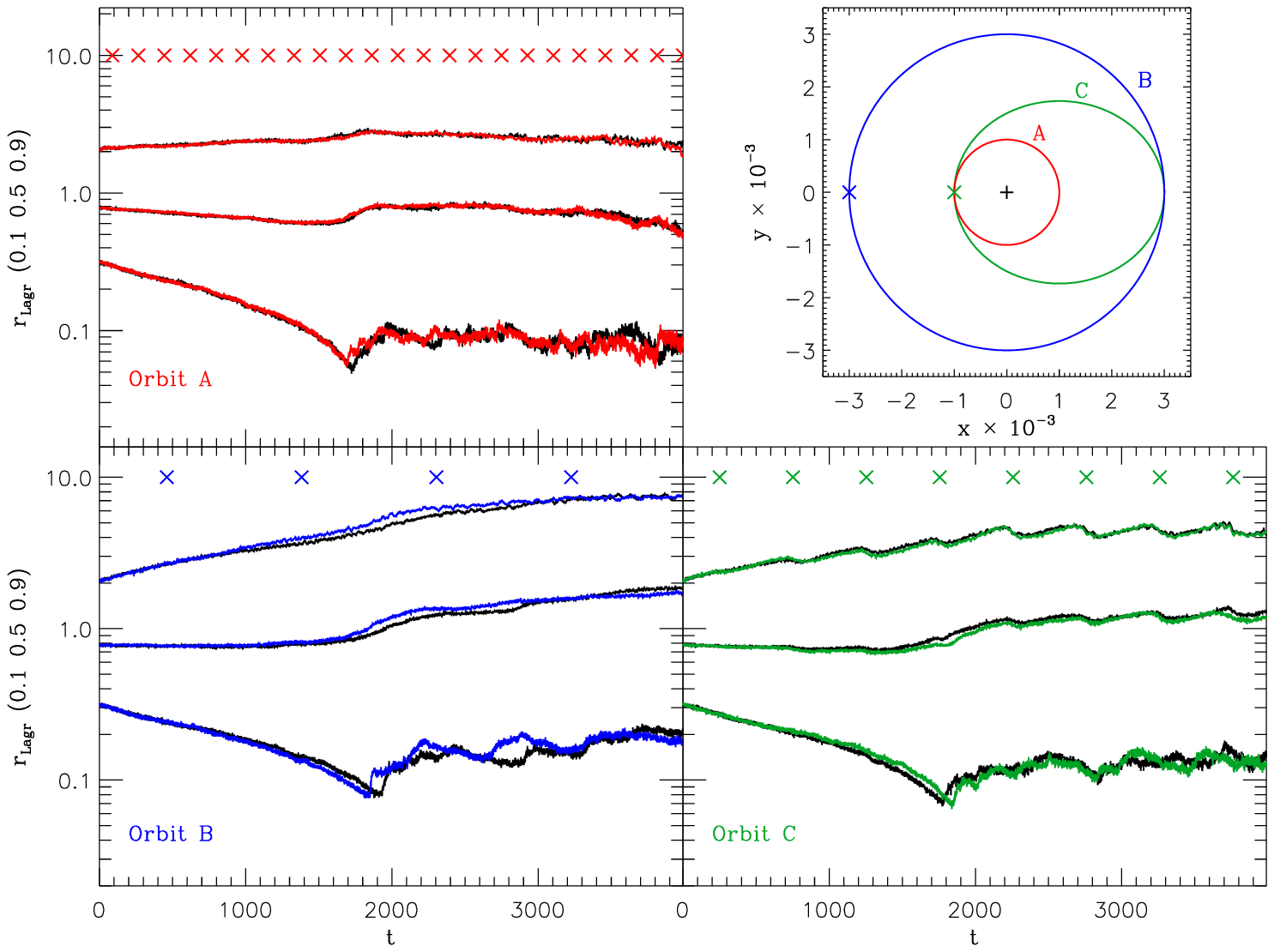}
\caption{Top-right panel: orbits of the clusters in the galactocentric frame. Other panels: Evolution of the 10\%, 50\% and 90\% Lagrange radii of the cluster on the three orbits, computed by the built-in method (black) and by \nbodytt (colour). On the top of each panel, crosses mark the passages of the cluster at its left-most position in the top-right panel (i.e. at pericentre for orbit C). The core-collapse phase is well visible at $t \sim 1700 - 2000$.}
\label{fig:lagrange}
\end{figure*}

To ensure the validity of the method, we ran a series of tests to compare the new implementation with the built-in methods\footnote{Option \#14 = 3 in \nbody.} of \nbody. For both codes, the centre of the cluster is fixed and the tidal field mimics the orbit of the galaxy around it. The orbits, shown in the top-right panel of \fig{lagrange}, are circular of radius $\rg=1000$ (A), circular of radius $\rg = 3000$ (B), and  elliptical with an eccentricity 0.5 and a pericentre of 1000 (C), which places the apocentre at a distance of 3000 so that the tidal field takes intermediate values, between the strong field (A) and the weaker one (B).

\begin{figure}
\includegraphics{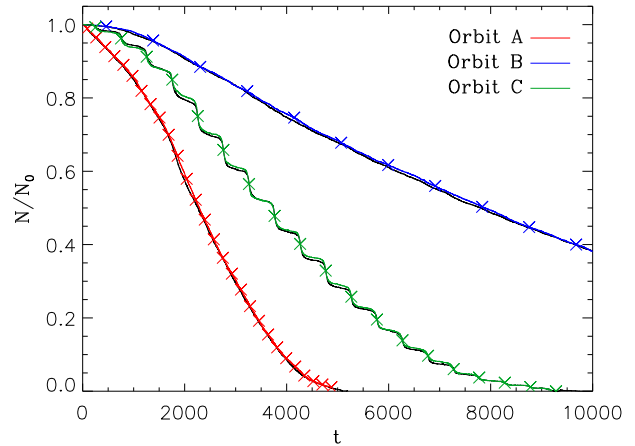}
\caption{Number of stars in the cluster along the orbits A, B and C around a point-mass galaxy, normalized to its initial value. The black curves represent the solution using the built-in method of \nbody, while the coloured ones are associated with the use of the tidal tensor. The symbols mark the passage of the cluster at the positions marked in the top-right panel of \fig{lagrange}.}
\label{fig:massloss_bivstt}
\end{figure}

The cluster is a \citet{1911MNRAS..71..460P} sphere made of $N=8000$ equal-mass particles. To focus on the effect of the tides, we have switched off stellar evolution. The parameters of the run are listed in \tab{units}. The evolution of some Lagrange radii of the cluster are plotted in \fig{lagrange}: before core-collapse, the relative differences ($1-r_\nbody/r_\nbodytt$) remains below 5\%, in all cases. The differences increases after core collapse, but not systematically, probably because of the formation of binaries which is sensible to numerical and $N$-body noises. As a complement, \fig{massloss_bivstt} plots the evolution of the number $N$ of stars in the cluster, normalized to its initial value. In both measurements ($r_\textrm{Lagr}$ and $N$), the agreement of the two approaches is very good for all the orbits. In particular along the elliptical orbit C, the expansion of the outermost layers of the cluster and the increased mass-loss near the pericentre passages is well-reproduced by the new method, both in term of time (epoch, delay and duration) and amplitude. These tests demonstrate that the interpolation scheme used to evaluate the tidal tensor at any time and the computation of the force done by \nbodytt allow us to retrieve the results obtained with well-tested methods, at a high level of accuracy.

\subsection{Other galactic profiles}
\label{sec:otherprofiles}

The method having been successfully tested, we now explore the evolution of clusters, still on circular orbits, but in the galactic potentials presented in Section~\ref{sec:circular} and \fig{potential}. The parameters $\mg$ and $\rhomu$ have been chosen so that the tidal radius is the same as that of the orbit A. The tensors are computed analytically along the orbit and passed to \nbodytt.

\begin{figure}
\includegraphics{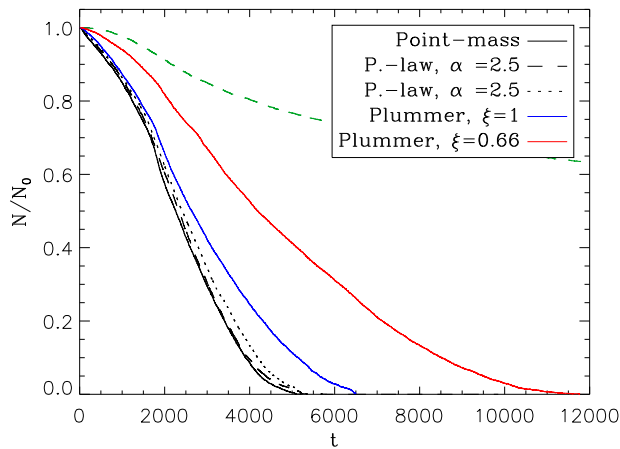}
\caption{Evolution of the normalized number of stars in the cluster on circular orbit in the galactic potentials presented in Section~\ref{sec:circular}, compared to the case of a cluster in isolation (dashed green line).}
\label{fig:massloss}
\end{figure}

The evolution of $N$ is presented in \fig{massloss}. The life-times of the clusters are ordered as predicted in Section~\ref{sec:escape}. We notice however that, for example, the life-time of the cluster in the Plummer galaxy at $\xi=0.66$ is $\approx 2.2$ time longer than that around the point-mass galaxy, while \eqn{dissolution} predicts a value of $\approx 1.57$. The discrepancy is due to a different evolution of the internal properties of the cluster ($\trh$, $C$, $\mc$), which are enclosed in the constant of proportionality of \eqn{dissolution}. Furthermore, because of the different shape of the potential (i.e. the second order effect in equation~\ref{eqn:dissolution}), the escape of stars does not occur at the same rate in all galaxies, which modifies $M_c$ and the tidal radius itself (i.e. the first order effect) differently from galaxy to galaxy. As a consequence, the effect due to the mild variation of the three-dimensional shape of the Jacobi surface gets strongly amplified with time.

Simulations at higher resolution ($N = 16\times 10^3$ and $32\times 10^3$ particles, but always the same $\mc$) have also been done, with these tidal fields. Not surprisingly, \nbodytt shows that the actual life-time of the clusters increases as $\sim N^{3/4}$ (see equation~\ref{eqn:baumgardt} with the usual scaling of $\trh$ with $N$), so that the discrepancy between the analytical values of $\tdiss$ and the life-times measured numerically for our five galactic cases is preserved at higher resolution. This demonstrates that our conclusions are not affected by the low-$N$ statistics of our experiments.

It is therefore very involved to derive analytically the mass or the life-time of clusters from initial parameters only. The use of numerical methods like \nbodytt provides a solution, but at the non-negligible cost of computational time.

\subsection{Fully arbitrary tidal field}
\label{sec:antennae}

As a last step toward generality, this Subsection presents the results obtained for a complex and highly time-dependent tidal field. The orbit chosen is extracted from a simulation of the Antennae galaxies (NGC~4038/39), a prototypical major merger. The galactic run, the parameters and the orbit are described in \citet[Fig. 8, orbit B, see also their Fig. 3]{2009ApJ...706...67R}: the cluster starts orbiting in the disc of NGC~4038 at $\sim 6 \kpc$ (on average) from the galactic centre; then it is ejected by the first galactic pericentre passage into the intergalactic bridges before falling back into the central region and remaining there for the rest of the merger. The Antennae being a real, observed object, the units are now scaled according to the galactic simulation (based on the observed spatial extension of the tidal tails and the peak radial velocity). The \nbodytt run is arbitrarily started $100 \Myr$ before the first pericentre passage of the two galaxies. The orbit of the cluster has also been integrated within its host galaxy (NGC~4038) in isolation, as a reference simulation. In both cases, the cluster is setup identically to those of the previous Sections (see Table~\ref{tab:units}). Our physical scaling makes it comparable in mass ($8000\msun$) and density ($\sim 2000\mpc$ within the half-mass radius) to Westerlund~1 or NGC~3603 \citep{2010ARA&A..48..431P}.

\begin{figure}
\includegraphics{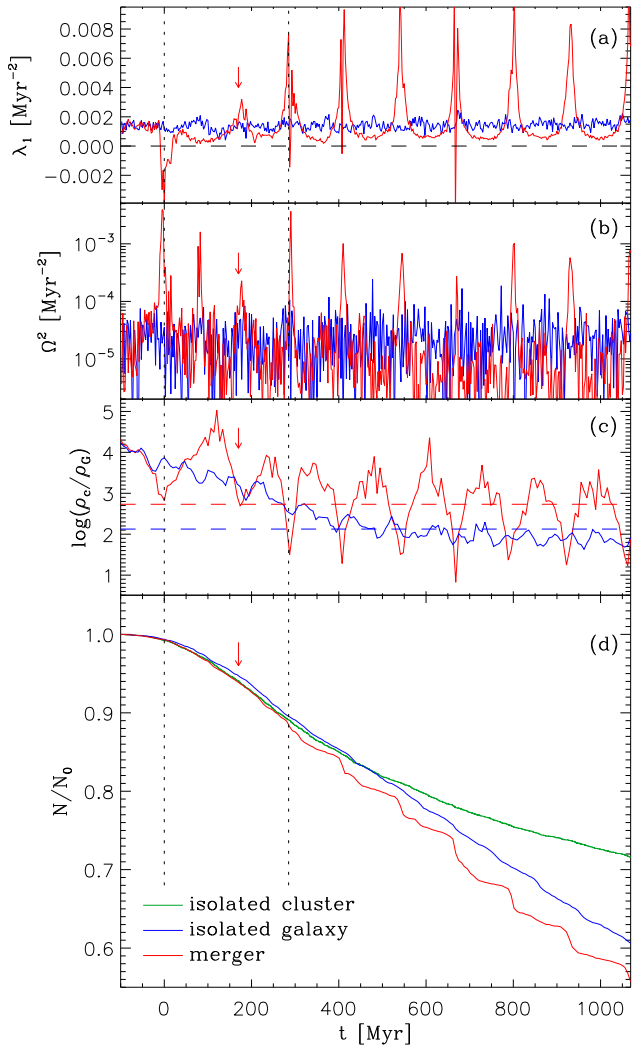}
\caption{(a): Maximum eigenvalue of the tidal tensor along the orbit in the Antennae (see text, red), and in the same progenitor galaxy but isolated (blue). (b): Instantaneous centrifugal contribution to the effective acceleration. (c): Logarithm of the ratio of the density of the cluster and the local density of the galaxy. The horizontal dashed lines show the ratios averaged over the life-time of the cluster. (d): Evolution of the normalized number of stars in the clusters in these galaxies, and in an isolated cluster (green). Vertical dotted lines mark the galactic pericentre passages and the arrow indicates when the cluster is in the bridges in the simulation of the Antennae.}
\label{fig:antennae}
\end{figure}

The maximum eigenvalue $\lambda_1$ of the tidal tensor is plotted in \fig{antennae}.a, in the case of the merger (red) and of the isolated galaxy (blue). The cluster is in compressive mode for $\lambda_1<0$. The centrifugal term\footnote{To evaluate the centrifugal term $\Omega^2$, we use the eigenvector $\bb{\nu}_1$ associated with $\lambda_1$. This vector points towards the main source of gravitation and thus, the variation of its direction gives the instantaneous orbital rotation speed. We obtain it via the dot product of two consecutive (normalized) $\bb{\nu}_1$'s:
\[
\Omega(t) \approx \frac{\textrm{acos}{\left[\bb{\nu}_1(t) \cdot \bb{\nu}_1(t-\dd t)\right]}}{\ddd t}.
\]} $\Omega^2$ is shown in logarithmic scale in \fig{antennae}.b. The peaks denote the velocity kicks the cluster receives when it is gravitationally slingshot. \fig{antennae}.c displays the ratio of the density of the cluster and the local density of the galaxy\footnote{The local density of the galaxy is given by the trace of the tidal tensor, through Poisson's law: 
\[
\sum_i \lambda_i = -\nabla^2 \phig(\br) = -4\pi G \rho_{\rm{G}}(\br).
\]}. Finally, the evolution of the number of stars in the cluster is displayed in \fig{antennae}.d, and compared to that of the cluster in isolation (green).

At $t\simeq 180 \Myr$, the cluster is in the extensive regime of the tidal bridges, marked by a slow increase of $\lambda_1$: the mass-loss accelerates with respect to that of the cluster in the isolated galaxy. During the second galactic pericentre passage ($t\simeq 285 \Myr$), the tidal field is mostly extensive, which once again, enhances the escape of stars. Later, the galaxies have merged and the cluster is orbiting the remnant. Its orbital period ($\sim 140 \Myr$) is clearly visible in both $\lambda_1$ and $\Omega^2$, but also in the mass-loss which is accelerated at pericentre, in a comparable fashion as the elliptical test-case of \fig{massloss_bivstt}. The rapid variations of $\lambda_1$ near the pericenter passages exist because the cluster flies at high speed in a highly asymmetric potential, which can be momentarily compressive at the location of the cluster.

Along this particular orbit, the local galactic density is, on average, about three orders of magnitude smaller than that of the cluster but the ratio of the two is peaked at several epochs. Therefore, the tidal field only affects the cluster during precise and short periods of time: the cluster is dense, robust enough to remain mildly disturbed on the long timescale. In other words, the life-time of this cluster is almost independent of the merger; only its `life-style' differs from the case of the isolated galaxy. However, the analysis of this single case is not statistically relevant to conclude that the merger has no secular impact on its clusters. The properties of the cluster population, in particular the cluster age function, rely on many parameters (structural and orbital) and we leave their study to a forthcoming paper.

\section{Summary, limitations and conclusions}

In this contribution, we propose a new formalism to describe the tidal field in $N$-body simulations by the means of tensors. Although the analytical approach rapidly becomes too involved for accelerated motions, we have derived the expressions of quantities representing the effect of the tides on stellar systems for circular orbits, with no restriction on the shape of the external potential. The main results of this study are:
\begin{itemize}
\item The use of the tidal and effective tensors allows us to simplify the representation of the problem, without loss of information (Section~\ref{sec:tensors}). Useful quantities like the tidal radius (equation~\ref{eqn:rt}), the energy of the Jacobi surface (equation~\ref{eqn:ej}) or the escape timescale (equation~\ref{eqn:escapetime}) can be easily computed for a given cluster.
\item The tidal radius is a first order approximation of the effect of a galaxy on a star cluster. The three-dimensional shape of the galactic potential has a second order effect (Section~\ref{sec:escape} and equation~\ref{eqn:dissolution}).
\item For a given tidal radius, a point-mass is the most efficient galactic profile in dissolving a cluster. Shallower potentials lead to longer dissolution times (\fig{dissolutiontime}).
\item A cluster in compressive tidal mode loses its stars more rapidly than in isolation, but significantly more slowly than if it was in extensive mode (Section~\ref{sec:escape}).
\item The knowledge of the evolution of the cluster parameters (half-mass radius, mass, energy distribution) is key to estimate the cluster life-time (Section~\ref{sec:escape} and Appendix~\ref{app:flux}).
\item For non-circular orbits, the analytical approach becomes very involved so that general and simple formulae do not exist (Section~\ref{sec:noncircular}).
\end{itemize}
To overcome this issue, we have developed and implemented a numerical method called \nbodytt which computes the evolution of $N$-body models of star clusters in any tidal field, by means of the pre-calculation of inertial tidal tensors. This method has been successfully tested and applied to innovative cases. Our main conclusions are:
\begin{itemize}
\item The tidal force felt by the stars of a star cluster can be accurately computed by evaluating the tidal tensor at all time by means of quadratic interpolation (Section~\ref{sec:tests}).
\item The dependence of the dissolution time on the three-dimensional shape of the galactic potential highlighted in Section~\ref{sec:escape} is confirmed by our numerical experiments. The numerical method does not suffer from the lack of knowledge of the evolution of the cluster properties that one has to face in a (semi-)analytical approach (Section~\ref{sec:otherprofiles}).
\item Our implementation has been applied to the complex case of a cluster in the major merger of the Antennae galaxies (Section~\ref{sec:antennae}). Its mass-loss reflects the nature of the time-dependent tidal field experienced along the orbit. In particular, the alternation of extensive and compressive tidal modes strongly affects the instantaneous dissolution rate, over a time-scale of several $10^7 \yr$.
\end{itemize}

This preliminary study shows the way to a very wide range of possible applications. However, one should keep in mind that our study is limited in several aspects. On the one hand, our cluster simulations are gas-free. Therefore, we cannot address the important point of the early life of the star cluster, prior to gas expulsion (first $\sim 10^6 \yr$ of the cluster's life), and we limit our study to initially relaxed systems. On the other hand, the second half of the galaxy-cluster coupling is not taken into account by our method: the feedback from stellar evolution, the escape of stars in the field and the formation of long tidal tails or streams are not implemented. Although they have a limited impact on the evolution of the cluster itself, it would be important to monitor their effects at larger scale (e.g. the metal enrichment of the interstellar medium). Furthermore, the mass-loss experienced by a cluster could affect its orbit within its host galaxy and thus, change the tidal field. In our approach, the scale decoupling does not allow to account for such effect.

To conclude, we have shown that a star cluster plunged in a time-dependent tidal field leads to a complex evolution, out-of-reach of (semi-)analytical approaches. A numerical method like the one proposed by \nbodytt provides a framework for future explorations of the role of the tides on star clusters. Among others, the use of a stellar mass function, primordial binaries and high order stars, stellar evolution and stellar mass-loss are as many lines of investigation to be pursued on top of the evolution in a background, time-dependent, galactic potential.

In a forthcoming paper, we will use more detailed galactic simulations including a prescription on the star formation, to compute the tidal tensors of an entire cluster population. \nbodytt will help us to derive the cluster- mass and age functions and their evolution, in various type of galaxies, in particular in mergers.

\section*{Acknowledgments}
We thank Douglas Heggie for interesting discussions that greatly helped to improve our understanding of the topic, and Jorge Pe\~narrubia, Dominique Aubert and the referee for their input. Sverre Aarseth is warmly acknowledged for making \nbody available, as well as Keigo Nitadori for the development of the GPU part of the code. FR thanks a visiting grant of the Institute of Astronomy of Cambridge where part of this work was done, as well as support from the EC through grant ERC-StG-257720; MG acknowledges the Royal Society for financial support.

\bibliographystyle{mn2e}
\bibliography{biblio}

\appendix

\section{Existence of the Lagrange points}
\label{app:compressive}

The Lagrange points $L_1$ and $L_2$, which define the tidal radius $\rt$, exist when the internal gravitational acceleration of the cluster is balanced by the effective tidal acceleration:
\begin{equation}
\frac{G\mc}{\rt^2} = \lambeone \rt.
\end{equation}
Along a circular orbit of radius $\rg$, one gets
\begin{equation}
\frac{G\mc}{\rt^2} = \left[-\left(\frac{\partial^2 \phig}{\partial r^2}\right)_{\rg} + \Omega^2 \right]\rt,
\end{equation}
which can be re-written by introducing the epicycle frequency $\kappa$:
\begin{equation}
\frac{G\mc}{\rt^2} = \left[-\left(\kappa^2-3\Omega^2\right) + \Omega^2 \right]\rt.
\end{equation}
This tells us that the tidal radius can be defined for $\kappa^2/\Omega^2 < 4$, which is always the case since the maximum value $\kappa^2/\Omega^2 = 4$ is reached for homogeneous mass distributions. To conclude, on a circular orbit, the centrifugal acceleration always compensates the tidal component, even in compressive mode, so that the Lagrange points $L_1$ and $L_2$ always exist.

\section{Computation of the escape time}
\label{app:flux}

In this Appendix, we generalize the expression of the escape time derived by \citet{2000MNRAS.318..753F} for a cluster in a circular orbit around a point-mass galaxy, to the case of any galactic potential. First, the origin of the coordinates is shifted to $L_1$ and the total potential is expanded to second order, so that
\begin{equation}
\phi-E_J = \frac{\lambeone}{2} \left[-3x^2 + y^2 + z^2\left(1-\frac{\lambethr}{\lambeone}\right) \right].
\end{equation}
The flux of phase volume across the new $x=0$ is expressed as
\begin{equation}
\mathcal{F}(E) = \int_{\dot{x}>0} \delta\left(\phi + \frac{v^2}{2} - E\right) \dot{x} \dd\dot{x}\dd\dot{y}\dd\dot{z}\dd y\dd z,
\label{eqn:flux}
\end{equation}
where the dot indicates derivation with respect to time and $\delta$ is the Dirac function. We change variables to $w:\dot{x} \mapsto \phi + v^2/2 -E$ so that $\ddd w = \dot{x}\dd\dot{x}$ and integrate the Dirac function over $w$ to get
\begin{equation}
\mathcal{F}(E) = \int_{\dot{x}>0} \dd\dot{y}\dd\dot{z}\dd y\dd z,
\end{equation}
with integration boundaries satisfying
\begin{equation}
2(E-\ej)-\lambeone \left[y^2 + z^2\left(1-\frac{\lambethr}{\lambeone}\right) \right] > 0.
\label{eqn:integrationboundaries}
\end{equation}
We change the remaining four variables into the hyper-spherical coordinates $\{R,\theta,\tau,\psi\}$, i.e.
\begin{equation}
\left\{\begin{array}{l}
y =  \lambeone^{-1/2}  R \cos{\theta}\\
z =  \left[\lambeone \left(1 - \frac{\lambethr}{\lambeone} \right) \right]^{-1/2} R \sin{\theta} \cos{\tau} \\
\dot{y} = R \sin{\theta} \sin{\tau} \cos{\psi}\\
\dot{z} = R \sin{\theta} \sin{\tau} \sin{\psi}
\end{array}\right. \textrm{with}\left\{
\begin{array}{l}
R > 0\\
\theta \in [0,2\pi] \\
\tau \in [0,\pi] \\
\psi \in [0,\pi]
\end{array}\right.
\end{equation}
so that the condition \eqn{integrationboundaries} becomes
\begin{equation}
2(E-\ej)-R^2 > 0.
\end{equation}
The determinant of the Jacobian matrix of the transformation gives the hyper-volume element:
\begin{equation}
\dd\dot{y}\dd\dot{z}\dd y\dd z = \frac{R^3 \sin^2{\theta} \sin{\tau}}{\lambeone\sqrt{1 - \frac{\lambethr}{\lambeone} }}\dd R\dd\theta\dd\tau\dd\psi.
\end{equation}
The flux is finally
\begin{equation}
\mathcal{F}(E) = \frac{2\pi^2 (E-E_J)^2}{\lambeone\sqrt{1 - \frac{\lambethr}{\lambeone}}}.
\label{eqn:fluxresult}
\end{equation}
The total flux is $2\mathcal{F}$ because stars can escape through apertures around two Lagrange points.

Similarly, the phase-space volume can be written
\begin{equation}
\mathcal{V} = \int \delta\left(\phi + \frac{v^2}{2} - E\right)\dd^3r\dd^3v.
\label{eqn:vflux}
\end{equation}
We first take out the angular part of the velocity and integrate the Dirac function over $v$:
\begin{equation}
\mathcal{V} = 4\pi \int \sqrt{2(E-\phi)}\dd^3r.
\label{eqn:bigv}
\end{equation}
Defining the dimensionless quantities
\begin{equation}
\left\{\begin{array}{ll}
\Psi^\star &= (E-\phi)\ \rt / (G\mc)\\
r^\star &= r/ \rt,
\end{array}\right.
\end{equation}
and substituting in \eqn{bigv} yields
\begin{equation}
\mathcal{V} = 4\pi \sqrt{2}\ (G\mc)^{1/2}\ \rt^{5/2}\ C,
\label{eqn:v_app}
\end{equation}
where
\begin{equation}
C = \int \sqrt{\Psi^\star}\dd^3r^\star
\end{equation}
is a dimensionless quantity which describes the intrinsic properties of the cluster. Instead of integrating over the entire solid angle, we note that the flux $\mathcal{V}$ is non-zero only at the vicinity of the Lagrange points $L_1$ and $L_2$. In a first approximation, we may consider a non-zero flux only at the exact position of $L_1$ and $L_2$. In that case, we replace the angular dependence of the previous integral with Dirac functions so that only two angular directions remain. That is, $C$ becomes
\begin{equation}
C = 2 \int \sqrt{\Psi^\star}\ {r^\star}^2\dd r^\star.
\end{equation}
For King profiles with the usual parameter $\Psi_0/\sigma^2$ ranging from 3 to 12, we found (by means of numerical integrations) that $C$ takes values ranging from 0.38 to 0.39 with a maximum reached for $\Psi_0/\sigma^2 \approx 8$.

By using the general definitions given in Section~\ref{sec:generalcase}, we find
\begin{equation}
\mathcal{V} = 4\pi \ C \sqrt{2}\ (GM_c)^{4/3}\lambeone^{-5/6}.
\label{eqn:v}
\end{equation}
It follows that the timescale for escape with the energy $E$ is
\begin{equation}
\tesc(E) = \frac{\mathcal{V}}{2\mathcal{F}(E)},
\end{equation}
which gives \eqn{escapetime}.

\end{document}